\documentclass[a4paper, notoc]{JHEP3}
\usepackage[dvips]{graphicx}
\usepackage{amsfonts}
\usepackage{amssymb}
\usepackage{epsfig}
\usepackage{cite}

\usepackage{amsfonts}
\usepackage{amsmath}
\usepackage{amssymb}

\usepackage{epsfig,graphicx,amssymb,amsfonts}

\def \beq {\begin{equation}}
\def \eeq {\end{equation}}
\def \bea {\begin{eqnarray}}
\def \eea {\end{eqnarray}}

 \def\e{{\rm e}}

\def\Z#1{_{\lower2pt\hbox{$\scriptstyle#1$}}}
\def\X#1{_{\lower2pt\hbox{$\scriptscriptstyle#1$}}}

 \def\sgn{\,\hbox{sgn}}

\def\IR{{\hbox{{\rm I}\kern-.2em\hbox{\rm R}}}}

\title{Simple cosmological de Sitter solutions on ${\rm dS}_4 \times {\rm Y}_6$ spaces}

\author{Ishwaree P. Neupane\\
Department of Physics and Astronomy, University of Canterbury\\
Private Bag 4800, Christchurch 8041, New Zealand
\\\email{E-mail:ishwaree.neupane@canterbury.ac.nz}}

\abstract{Explicit time-dependent solutions of the 10D vacuum
Einstein equations are found for which spacetime is compactified
on six-dimensional warped spaces. We explicitly work out an
example where the internal manifold is a six-dimensional
generalized space having positive, negative or zero scalar
curvature, whose base can be a five-sphere $S^5$ or an Einstein
space $T^{1,1}=(S^2\times S^2)\rtimes S^1$. In this paper,
inflationary de Sitter solutions are found just by solving the 10D
vacuum Einstein equations. Our results further show that the
limitation with warped models studied to date has arisen partly
from an oversimplification of the 10D metric ansatz. We also give
some explicit examples of a
non-singular warped compactification on de Sitter space dS$_4$.\\

  }


\preprint{UOC-TP 011/09, ~ arXiv:0901.2568}

\begin{document}

\section{Introduction} The observed current acceleration of our
universe~\cite{supernovae} is among the most puzzling discoveries
in modern cosmology. This somewhat unexpected result together with
an inflationary epoch required to solve the horizon and flatness
problems of the big bang cosmology need to be understood in the
framework of fundamental theories.
While the basic principles of the early inflation are rather well
established, with many of its predictions being supported by
observational data from WMAP~\cite{WMAP}, it still remains a
paradigm in search of a concrete theoretical model. Efforts are
still underway to explain accelerating universes from string
theory -- a consistent quantum theory of gravity in $9+1$
spacetime dimensions.

In light of observational evidence for both an inflationary epoch
in the distant past and a recent cosmic acceleration, it is of
importance to construct explicit de Sitter solutions from 10D or
11d supergravity models which are the low-energy effective
theories of superstrings. In recent years, various low energy
versions of string theory have been widely used to understand and
to probe the mathematical structures of quantum field theories and
cosmologies at a microscopic scale. String theory has been a
source of new ideas and has greatly inspired novel scenarios of
cosmology, but it has yet to confront data and make predictions.
Examples of predictions which have received considerable interests
include mechanism of supersymmetry breaking, techniques of
generating a small positive cosmological constant~\cite{KKLT} or
quintessence~\cite{Ish03c} and general statements about the scale
and properties of inflation~\cite{Becker:2007}. Motivated by this
success and observation driven cosmology, it is natural to study
string theory models in cosmological
backgrounds~\cite{Chen:02yq,Ohta:2003pu,Ish03a,Ish05a,Ish07a}.

The first requirement for any predictable model based on string
theory (or theories of higher dimensional gravity) is to find a
mechanism of dimensional reduction from $10$ (or $11$) dimensions
to $4$. The simplest scenario of this type is ordinary
Kaluza-Klein compactification of a 10D supergravity theory with
geometry of the form $M \equiv M_{4}\times {\cal M}_6$, where
$M_{4}$ is a four dimensional space-time with Lorenzian signature
and ${\cal M}_6$ is a six-dimensional compact manifold. Models in
four dimensions obtained through ordinary KK reduction on a
Ricci-flat space (or string theory compactified on Ricci flat
Calabi-Yau spaces) are, however, far from being phenomenologically
interesting as they are plagued at the classical level by a
plethora of massless scalar fields $\phi_i$, which are not
observed in our real world. A viable 4D effective theory, like
Einstein's theory of general relativity, must feature a
cosmological constant-like term or a nontrivial scalar field
potential. The latter could on the one hand lift the moduli
degeneracy, thereby making the theory predictive, and on the other
hand select a vacuum state for our universe with some desirable
properties, such as a small cosmological vacuum energy or
gravitational dark energy.

An important problem in string or higher dimensional cosmology is
to explain the asymmetry between the sizes of large and small
spatial dimensions. A reasonable explanation for the observed
asymmetry would be a cosmological evolution in which all
dimensions are initially small (and symmetric) and are then
subject to asymmetric expansion driven by the underlying
cosmological dynamics. It would be interesting to see if
inflationary cosmologies and/or the origin of three large spatial
dimensions can be understood as purely an outcome of the vacuum
Einstein equations in a cosmological (time-dependent) background.

To construct a phenomenologically viable model, one could either
consider a compact internal space and then deal properly with the
stabilization of the extra dimensional volume, or else consider a
noncompact solution and then introduce a method of inducing
finite-strength four dimensional gravity. Particularly, in the
context of ten-dimensional warped supergravity models, some of the
extra dimensions could extend along the physical 4D hypersurface.
In this scenario the effective 4D Newton's constant can be finite
because of a strong warping of extra dimensions as in the
Randall-Sundrum braneworld models~\cite{RS}.

In this paper, we report on new cosmological solutions that
explain inflationary universes utilizing exact solutions of 10D
Einstein equations. These solutions correspond to the dimensional
reduction to 4 dimensions of the type II supergravity, where the
spacetime is a warped product of a six-dimensional generalized
space $Y_6$ and $M_4$ ($\equiv \IR^{1,3}$). We also outline
possible observational consequences of these new exact solutions,
which may have significant contribution in current efforts to make
contact with cosmological observations.

\section{Ricci-flat spaces and time-dependent solutions}

Assuming that ten-dimensional supergravity is the relevant
framework, we start with the following 10D metric ansatz (in
Einstein-conformal frame):
\begin{equation}\label{10D-metric-gen}
ds^2\Z{10} = e^{-6\varphi(x)} e^{2A(y)}  \hat{g}_{\mu\nu} dx^\mu
dx^\nu +e^{2\varphi(x)} e^{- \alpha A(y)}  {g}_{mn}(y) dy^m dy^n,
\end{equation}
where $e^{\varphi(x)}$ is the Weyl rescaling factor, $A(y)$ is the
warp factor as a function of one of the internal coordinates, $y$,
and $\alpha$ is a constant. For the present study, which will be
based on 10D vacuum Einstein equations, the above metric ansatz is
sufficiently general. The metric of the usual $3+1$ spacetime (or
a Friedmann-Robertson-Walker universe) is
\begin{equation}\label{FLRW}
ds^2\Z{4} \equiv \hat{g}_{\mu\nu} dx^\mu dx^\nu \equiv -
a^{2\delta} du^2+ a^2 d{\bf x}^2,
\end{equation}
where $a(u)$ is a scale factor and $\delta$ is a constant. The
cosmic time $t$ is defined by $dt=a^\delta du$. The metric of the
internal six-dimensional space $Y_6$ may be written as
\begin{equation}\label{6D-met-ansz}
ds^2\Z{6} = {g}_{mn}(y) dy^m dy^n = dy^2+ f(y) ds^2\Z{X_5}.
\end{equation}
The 5D base space $X_5$ may be taken to be an Einstein space
$T^{1,1}= (S^2\times S^2)\rtimes S^1$ with the metric
\begin{equation}\label{simple-coni}
ds^2\Z{X_5}= g_{ij}d\Theta^i d\Theta^j=\frac{1}{9}\, e_\psi^2 +
\frac{1}{6} \left(e_{\theta_1}^2+e_{\phi_1}^2
+e_{\theta_2}^2+e_{\phi_2}^2\right),
\end{equation}
where $e_{\theta_i}=d\theta_i$, $e_{\phi_i}=\sin\theta_i d\phi_i$,
$e_\psi\equiv d\psi +\cos\theta_1 d\phi_1+\cos\theta_2 d\phi_2$ or
some other compact manifolds, such as $S^5$. In the above,
$(\theta\Z1,\phi\Z1)$ and $(\theta\Z2, \phi\Z2)$ are coordinates
on each $S^2$ and $\psi$ is the coordinate of a $U(1)$ fiber over
the two spheres. In this particular example, the internal space
$Y_6$ becomes Ricci flat ($R(\tilde{g})=0$) when $f(y)\equiv y^2$.
The metric~(\ref{6D-met-ansz}) then defines a standard 6D conifold
widely studied in the literature, in the context of warped
supergravity solutions; see, for example~\cite{KT}.

Before presenting some explicit solutions, we shall make some
general comments about our choice of the 10D metric ansatz. In
dimensions $D\ge 6$, and with $A(y)\ne {\rm const}$, one cannot
absorb the factor $e^{-\alpha A(y)}$ into $g_{mn}(y)$ just by
using some coordinate transformations unless that each and every
components of the metric $g_{mn}(y)$ are equal or proportional to
the same function, say $f(y)$. For brevity, let us take $D=10$ and
write the 6D metric as
\begin{equation}
ds\Z{6}^2 = h(y)\,dy^2 + f(y)\, \tilde{g}_{mn} d\Theta^m
d\Theta^n,
\end{equation}
where $\tilde{g}_{mn}$ denote the metric components of the base
space $X_5$, which are independent of the $y$ coordinate. In order
to absorb the factor $e^{-\alpha A}$ inside $dy^2$ one could use
the relation $e^{-\alpha A(y)} h(y) dy^2 \equiv d\tilde{y}^2$ and
also define a new function $X(\tilde{y})$ such that $e^{-\alpha
A(y)} f(y) \equiv X(\tilde{y})$. The warp factor $e^{2A(y)}$
multiplying the 4D part of the metric is now
$[X(\tilde{y})/f(\tilde{y})]^{-2/\alpha}$. As is evident, the 10D
metric still involves two unknown functions and the free parameter
$\alpha$. Thus, in order to solve 10D Einstein equations in full
generality, we should perhaps allow in the 10D metric ansatz one
more free parameter, like $\alpha$ in the above example, than
those considered in the literature, for
example,~\cite{Gibbons-84,Malda:2000,GKP}.

It is easy check that, with the metric (\ref{10D-metric-gen}), the
10D vacuum Einstein equations allow explicit solutions only when
$\frac{dA}{dy}\frac{d\varphi}{dt}=0$, which is the vanishing
condition of the Ricci tensor $R_{ty}$. In this paper we only
study explicit solutions of 10D vacuum Einstein equations, so we
may consistently demand that either $A(y)={\rm const}$ or
$\varphi(t)={\rm const}$. Of course, these conditions can be
relaxed by introducing more complicated metric fluxes (or
geometric twists) or by introducing external fluxes (or form
fields) as present in various versions of supergravity or string
theory. Alternatively, as in~\cite{Ish05a,Ish07a}, one could allow
time-varying volume factors to each of the factor spaces, such as
$X_5= (S^2 \times S^2)\rtimes S^1$.

First, let us assume that the Weyl factor $e^{2\varphi}$ is time
varying, $d\varphi/dt\ne 0$. In the vacuum case, the condition
$R_{ty}=0$ implies that $A(y)= {\rm const}$. In this case the
exact solution to 10D Einstein equations following from equations
(\ref{10D-metric-gen})-(\ref{simple-coni}) is given by
\begin{equation}
a(u)\propto e^{c\Z0 u}, \qquad \varphi=\pm \frac{1}{2} \ln a,
\qquad \delta=3.
\end{equation}
where $c\Z0$ is arbitrary. The same result can be obtained by
considering more general Ricci flat spaces, such as
\begin{equation}\label{gen-6D-metric}
ds^2_6= f\Z1(y) \,dy^2+ f\Z2(y)\, ds^2\Z{X_5}.
\end{equation}
The Ricci-flatness condition ${R}_6=0$ implies that $4 f\Z1 f\Z2 =
(df\Z2 /dy)^2$, which is satisfied, for instance, with $f\Z1 (y)
\equiv f\Z0$ and $f\Z2(y) =f\Z0 (y+y\Z0)^2$. A more general 6D
metric ansatz is
\begin{eqnarray}\label{gen-res-coni}
ds\Z{6}^2 = \lambda^2 f\Z1(y)\,dy^2 + \frac{\lambda\Z{1}^2}{9}
f\Z2(y) y^2 e_\psi^2 + \frac{\lambda\Z{2}^2}{6} y^2
\left(e_{\theta_1}^2+e_{\phi_1}^2\right) +
\frac{\lambda\Z{3}^2}{6} (y^2+6 b^2)
\left(e_{\theta_2}^2+e_{\phi_2}^2\right),
\end{eqnarray}
where $\lambda$, $\lambda\Z{i}$ are arbitrary constants and $b$ is
the resolution parameter. This metric becomes Ricci flat when
\begin{equation}
f\Z{1}^{-1}(y)=f\Z{2}(y)=\frac{y^6+9 b^2 y^4 +c}{y^4(y^2+6 b^2)},
\end{equation}
where $c$ is an arbitrary constant and $\lambda\Z1
=\lambda\Z2=\lambda\Z3=\lambda$. In this case the
metric~(\ref{gen-res-coni}) defines a standard 6D resolved
conifold considered previously by Tseytlin {\it et
al}~\cite{Zayas:2000}. The importance of taking a warped conifold
metric has already been emphasized in the literature,
see~\cite{Ish07a} for a discussion in a time-dependent
(cosmological) background and~\cite{KT,KS,Zayas:2000} for some
discussions in a static supergravity background.

In the case of $A(y)={\rm const}$ and a Ricci-flat 6D space
(irrespective of its topology), the solution to 10D vacuum
Einstein equations is given by
\begin{equation} a \propto (t+t\Z1)^{1/3},
\qquad \varphi=\varphi\Z0 \pm \frac{1}{6}\ln (t+t\Z1).
\end{equation}
This gives only a non-accelerating universe. A more interesting
solution can be found by replacing the FRW metric (\ref{FLRW}) by
the metric of an open universe, in which case one has the
following critical solution:
\begin{equation}
a(t)=t+t\Z1, \qquad \varphi(t) ={\rm const}, \qquad A(y)={\rm
const}.
\end{equation}
A detailed analysis shows that one has to consistently set
$\varphi={\rm const}$ to get an inflationary de Sitter solution
from 10D Einstein equations, at least, in pure supergravity
models.

\section{Generalized 6D spaces and exact de Sitter solutions}

In this section we present a few explicit models of warped
compactification for which inflationary de Sitter solutions are
possible even in the vacuum case. These examples belong to
standard warped compactifications for which the external fluxes
are turned off or absent.

First we write the 10D metric in the following form (with
$\alpha=0$):
\begin{eqnarray}\label{gen-coni2}
ds_{10}^2 &=& e^{2A(y)} ds\Z{4}^2+  \lambda^2 \left(dy^2+
\frac{y^2}{2} \,ds\Z{X_5}^2\right).
\end{eqnarray}
One could in principle start with the metric of a generalized 6D
space given in (\ref{gen-res-coni}), but with $\alpha=0$ the 10D
vacuum Einstein equations admit an exact solution only when
$\lambda\Z1=\lambda\Z2=\lambda\Z3=\lambda/\sqrt{2}$ and $b^2=c=0$,
for which the metric (\ref{gen-res-coni}) reduces to the 6D part
of the metric(\ref{gen-coni2}). The explicit solution to 10D
Einstein equations is given by
\begin{equation}\label{warp-Omega-sol}
A(y) = \frac{1}{2}\ln\left(\frac{3 y^2}{8 L^2}\right), \qquad
a(t)\propto e^{t/{\lambda L}},
\end{equation}
where $L$ is arbitrary. Interestingly enough, a warped spacetime
as in (\ref{gen-coni2}) is also motivated from particle physics
consideration in the manner of the Randall-Sundrum
model~\cite{RS}, where one ignores the $X_5$ part of the 10D
metric. There is however a crucial difference from that in RS
models. In our case, the usual 4D spacetime and the internal 6D
space both have positive curvature: $\hat{R}_{(4)}= 6(\ddot{a}/a+
\dot{a}^2/a^2)= 12/(\lambda^2 L^2)$ and $\tilde{R}_{(6)}=
20/(\lambda^2 y^2)$. The 10D warped geometry is dS$_4\ltimes Y_6$.
It is readily checked that, for the
solution~(\ref{warp-Omega-sol}), the 10D Ricci scalar curvature
vanishes, $R_{10}=e^{-2A} \hat{R}_4+ \tilde{R}_6 -(4/ \lambda^{2})
(2A^{\prime\prime}+5 {A^\prime}^2 + 10A^\prime/y)=0$.

\medskip
Next, we write the 10D metric in the following form (with
$\alpha=2$)
\begin{equation}\label{10D-metric}
ds_{10}^2 =  e^{2A(y)}\,\hat{g}_{\mu\nu} dx^\mu dx^\nu +
e^{-2A(y)} \,ds^2_6.
\end{equation}
The 10D Einstein equations admit an exact solution when the 6D
metric takes the form
\begin{equation}\label{gen-conifold2}
ds^2_6= \lambda^2 \left( dy^2 + 2 y^2\, ds^2\Z{X_5}\right),
\end{equation}
which is actually obtained by taking
$\lambda\Z1=\lambda\Z2=\lambda\Z3= \sqrt{2}\, \lambda$ and
$b^2=c=0$ in (\ref{gen-res-coni}). In terms of the $u$-coordinate
(i.e. with $\delta=3$ in equation~(\ref{FLRW})), the explicit
solution is given by
\begin{equation}\label{GKP-case}
A(y)= \frac{1}{4}\ln \left(\frac{3 y^2}{2 L^2}\right), \qquad
a(u)= \left(\frac{\lambda^2 L^2}{9 u^2}\right)^{1/6}.
\end{equation}
Note that the singularity of the scale factor at $u=0$ is not a
singularity of the usual 4D spacetime. This may be understood by
looking at the same solution but expressed in terms of the 4D
proper time $t$, which is nothing but $a(t) \propto \exp \left(t /
\lambda L\right)$.

The above results clearly show that an accelerated expansion of
the non-compact directions require nothing extraordinary other
than a 6D Einstein space $Y_6$ (which may be positively curved as
in (\ref{gen-coni2}) or negatively curved as in
(\ref{gen-conifold2})) that supports not only a nontrivial warp
factor but also a nonzero Hubble parameter. To our knowledge, this
gives the first example where de Sitter solutions are found just
by solving 10D vacuum Einstein equations, i.e., without
introducing external fluxes or form fields.

Inflationary de Sitter solutions can be obtained also from warped
compactifications on some Ricci flat 6D spaces, provided we write
the 10D metric in a more general form. For illustration, let us
make the following ansatz
\begin{equation}\label{gen-coni-4}
ds^2_{10}= \e^{2A(y)} ds^2_4 + \e^{-\,\alpha A(y)} \lambda ^2
\left( dy^2+\alpha\Z{1}\, y^2 ds^2\Z{X_5}\right),
\end{equation}
where $\alpha$ and $\alpha\Z{1}$ are some constants. For this
metric ansatz, we get
\begin{subequations}
\begin{align}
\sqrt{-g\Z{10}} = \frac{\lambda ^6}{108}\, \alpha\Z{1}^{5/2} y^5
\e^{(4-3\alpha) A} \sin\theta\Z1\sin\theta\Z2
\,\sqrt{-\hat{g}\Z{4}}, \\
{\cal R}\Z{10}= \e^{-2A} \left(\hat{R}\Z{4}-{\cal L}\Z\Lambda
\right),
\end{align}
\end{subequations}
where $\hat{R}\Z{4} =6(\ddot{a}/a+\dot{a}^2/a^2)$ and
\begin{eqnarray}\label{4d-CC}
{\cal L}\Z{\Lambda} = \frac{\e^{(\alpha+2)A}}{\lambda ^2} \Big[
(8-5\alpha) \left(A^{\prime\prime}+\frac{5}{y} A^\prime \right) +
(20-16\alpha+5 \alpha^2){A^\prime}^2 +\frac{20}{y^2}
\frac{(\alpha\Z{1}-1)}{\alpha\Z{1}} \Big],
\end{eqnarray}
where $A^\prime=dA/dy$. The 6D scalar curvature is
$R_{(6)}=20(1-\alpha\Z{1})/(\alpha\Z{1} y^2)$. The effective 4D
theory can have a nonzero cosmological constant-like term even
when the 6D space is Ricci flat. This corresponds to the choice
$\alpha\Z{1}=1$. To be precise, we can explicitly solve the 10D
vacuum Einstein equations following from (\ref{gen-coni-4}), whose
solution is given by
\begin{eqnarray}\label{most-gen-sol}
\e^{(\alpha+2)A} = \frac{3(\alpha+2)^2 y^2}{32 L^2},\qquad
\alpha\Z{1}= \frac{(\alpha+2)^2}{8}, \qquad a(t) &\propto &
e^{t/{\lambda L}}.
\end{eqnarray}
Note that this solution will not change if we take $X_5= S^5$
instead of $X_5=T^{1,1}$. In the case of a Ricci-flat 6D space, we
shall take $\alpha= -2\pm 2\sqrt{2}$. Hence
\begin{equation}
e^{A(y)}= \left(\frac{3 y^2}{4L^2}\right)^{\pm 1/2\sqrt{2}},
\qquad \alpha\Z{1}=1.
\end{equation}
One takes the $+$ve exponent so as to avoid the divergence of the
warp factor at $y=0$.

In the context of string flux compactifications, Giddings {\it et
al}~\cite{GKP} made the choice $\alpha=2$ and $R\Z{6}=0$. This
already belongs to a restrictive class of metrics, for which the
solution of 10D vacuum Einstein equations is trivial: $a(t)={\rm
const}$, $A(y)={\rm const}$. In fact, as is clear from
(\ref{most-gen-sol}), when $\alpha=2$, an inflationary de Sitter
solution can be obtained by allowing $Y_6$ to have negative
curvature ${R}_6={R}_{mn}{g}^{mn}<0$, or by taking $\alpha\Z{1}=2$
in~(\ref{gen-coni-4}). Similarly, the model studied by Gibbons
in~\cite{Gibbons-84} corresponds, in our case, to the choice
$e^{A(y)}=W(y)$ and $\alpha=0$. In this case, cosmological de
Sitter solutions can be obtained by allowing $Y_6$ to have
positive curvature ${R}_6>0$, or by taking $\alpha\Z{1}=1/2$
in~(\ref{gen-coni-4}).

Note that the solutions given above,
equations~(\ref{warp-Omega-sol}) (\ref{GKP-case}) and
(\ref{most-gen-sol}), are defined up to the rescaling $y\to
y+\epsilon$, where $\epsilon$ is a shift parameter denoting the
minimum size of $X_5$. The 10D Kretschmann scalar is given by
$K=R_{abcd}R^{abcd}\propto (y+\epsilon)^{-8/(2+\alpha)}$. With
$\epsilon>0$, the metric is smooth in the range $0\le y \le
\infty$. The only singularity is such that the warp factor
$e^{2A}$ going to zero at $y=-\epsilon$. However, since the
$g_{00}$ component of the metric decreases as $y\to -\epsilon$,
according to the strong version of singularity theorem discussed
in~\cite{Malda:2000}, a space-like singularity where $e^{2A}\to 0$
may be allowed as it can have a field theory interpretation. In
any case, we can completely get rid of this pathology, i.e. a
causal treatment of bulk singularity, by expressing our solutions
in some more suitable coordinate system.

In fact, for the metric~(\ref{gen-coni-4}), the singularity of the
warp factor at $y=0$ (with $\alpha\Z{1}\ne 1$) is a coordinate
artifact. To quantify this, let us first consider the case
$\alpha=-2$ and introduce a new coordinate $z$ such that $y\equiv
e^{-\, \beta z}$, where $\beta$ is a constant. The ranges of the
coordinates are $- \infty \le z \le \infty$ and $0 \le y \le
\infty$. We then find that the metric
\begin{equation}
ds_{10}^2 = e^{2A(z)} \left(-dt^2 +a(t)^2 \,d{\bf x}^2\right) +2
e^{-2A(z)} r_c^2\, e^{-\,2 \beta z} \left(\frac{\beta^2}{2}
\,dz^2+ \frac{1}{9}\, e\Z{\psi}^2 + \frac{1}{6} \sum_{i=1}^{4}
e_i^2\right) \label{asymptote1}
\end{equation}
solves all of the 10D Einstein equations when
\begin{equation}
A(z) = - \frac{\beta(z+z\Z{0})}{2},  \qquad a(t)\propto \exp
\sqrt{\frac{2\,t^2\,e^{-\,2\beta z\Z{0}}}{3
\,r_c^2}},\label{asymptote3}
\end{equation}
where $r_c\equiv \lambda$ is the compactification scale. From the
relation $e^{-2A}= e^{2A} e^{2\beta(|z|+z\Z{0})}$, we can easily
see that it is the same warp factor $e^{2A}$ that multiplies both
the 4D and the 6D metrics in (\ref{asymptote1}). The Kretschmann
scalars for the 6D and 10D parts of the metric are given by
$R_{mnpq} R^{mnpq}=53\, e^{2\beta (z-z_0)}/(2 r_c^4)>0$ and
$R_{ABCD}R^{ABCD}= 107\, e^{2\beta(z-z_0)}/(3 r_c^4)>0$, which are
smooth everywhere, and the solution has no any physical
singularities.

In the $\alpha\ne -2$ case, using the transformation $y\equiv
e^{-z/2}$ in~(\ref{gen-coni-4}), we find that the explicit
solution to 10D Einstein equations is given by
\begin{equation}
A=-\frac{(z+z_1)}{2+\alpha}, \qquad a\propto \exp
\sqrt{\frac{32\,t^2 e^{-z_1}}{3(2+\alpha)^2 \lambda^2}}, \qquad
\alpha\Z{1}=\frac{(2+\alpha)^2}{8}.
\end{equation}
This solution is also regular everywhere.

Finally, we give one more explicit example of a completely
nonsingular solution. We write the 10D metric in the form
\begin{eqnarray}\label{10D-nonsin}
&& ds\Z{10}^2 = e^{2A(z)} ds\Z{4}^2 + e^{- \alpha A(z)} \lambda^2
\,ds_6^2,
\end{eqnarray}
where the 6D metric has the form
\begin{equation}\label{6D-nonsin}
ds^2_6 \equiv \sinh^2(z+z_0) \,dz^2+ \alpha\Z{1}\,\cosh^2(z+z_0)\,
ds^2\Z{X_5}.
\end{equation}
Both the 6D Ricci scalar and the Kretschmann scalar
$K=R_{mnpq}R^{mnpq}$, i.e.
\begin{equation}\label{Kretschmann}
{R}_6=\frac{20(1-\alpha\Z{1})}{\alpha\Z{1}\,\cosh^2{(z+z_0)}},
\qquad K=\frac{8(5\alpha\Z{1}^2-10\alpha\Z{1}+17)}{\alpha\Z{1}^2
\cosh^4 (z+z_0)},
\end{equation}
are smooth everywhere. The 10D Einstein equations are solved for
\begin{eqnarray}\label{nonsin-sols}
A(z)= \frac{1}{2+\alpha}\ln \left(\frac{3(2+\alpha)^2
\cosh^2(z+z_0)}{32 L^2}\right), ~
\alpha\Z{1}=\frac{(\alpha+2)^2}{8}, ~ a(t)&\propto &
e^{t/{\lambda L}}.\nonumber \\
\end{eqnarray}
This provides an explicit example of non-singular warped
compactification on dS$_4$. The above results also reveal that
inflationary cosmology is possible with every choice of $\alpha$,
provided that we do not restrict the spatial curvature of the
internal space.

\subsection{Einstein's equations on a brane}

In the following we specify a boundary condition such that the
warp factor is regular at $z=0$ where we place a p-brane with
brane tension $\tau_p$. We also impose a $Z_2$ symmetry around the
brane's position at $z=0$. The metrics like (\ref{asymptote1}) and
(\ref{10D-nonsin}) will then have discontinuity in its first
derivatives at $z=0$, implying that $\partial |z|/\partial
z=\text{sgn}(z)$. For illustration, we consider the metric
(\ref{asymptote1}) from which we derive
\begin{eqnarray}\label{10D-soln}
G_{zz}=0, \qquad G_{MN}= - \frac{8}{\beta r_c^2}\, e^{- \beta
z_0}\,\delta(z) g_{MN},
\end{eqnarray}
where $(M,N)=t,x_i, \theta_i, \phi_i, \psi$. The 10D Einstein
equations can be written as
\begin{equation}\label{10D-boundary}
G_{AB}=  - \tau_p\,P(g_{AB}),
\end{equation}
where $P(g_{MN})$ is the pull-back of the spacetime to the world
volume of the $p$-brane (with $p\ge 3$) with tension $\tau_p$.

In general, in a 10D warped background, one can consider a p-brane
(with co-dimensions $3\le p\le 8$) wrapped on the $X_5$ space of
the internal manifold. For instance, in the $p=8$ case, three of
the extra dimensions could extend along the $x^\mu$ directions. A
situation like this could arise in the type IIA supergravity in
ten dimensions where one finds NS5-branes as well as D6 and
D8-branes. Similarly, in the type IIB supergravity in ten
dimensions, one may consider both D3 and D5-branes.

Let us consider the special case that $p=8$, for which $P(g_{AB})=
\delta(z) \delta_A^M \delta_B^N\,g_{MN}/\sqrt{g_{zz}}$. From
equations~(\ref{10D-soln})-(\ref{10D-boundary}) we then find that,
with
\begin{equation}
\tau_p = \frac{8}{r_c}\,e^{-\beta z\Z{0}/2},
\end{equation}
the 10D Einstein equations are satisfied both at $z=0$ and away
from $z=0$.

\subsection{Some remarks on earlier no-go theorems}

A couple of remarks are in order. The no-go theorems discussed
earlier by Gibbons~\cite{Gibbons-84} and Maldacena and
Nu\~nez~\cite{Malda:2000} do not apply here, at least, in their
original forms. The main reason for this is that the theorems
discussed in~\cite{Gibbons-84,Malda:2000,Wesley} required all of
the extra dimensions to be physically compact, the warp factor to
be constant and the warped volume $V^{\text w}\Z{6} \sim \int
\sqrt{g\Z{6}} \,e^{(2-3\alpha)A}$ to be finite (or constant) when
integrated over $Y_6$. In our examples the extra-dimensional
volume is only suppressed as compared to the 4D volume, but it can
be arbitrarily large. The previous authors also imposed an extra
constraint, so-called a boundedness condition $\int \nabla_z^2
e^{n A}=0$ (for any positive $n$), which is generally not
satisfied by cosmological solutions, especially, in the presence
of localized sources like branes. In the following we would like
to be a bit more explicit on these remarks.

The discussions of the no-go theorem
in~\cite{Gibbons-84,Malda:2000,GKP} appeared to be special at
least from two reasons. First, the previous authors assumed,
though only implicitly, that the extra dimensional manifold is
maximally symmetric, such as (with $D=10$, $m=6$),
\begin{eqnarray}
ds\Z{6}^2 &=& dz^2+ dz_1^2+ \cdots + dz_{5}^2,\quad
(\epsilon=0, ~T^6)\label{6D-flat}\nonumber\\
ds\Z{6}^2 &=&  dz^2+ \sin^2{z}
\,d\Omega_{5}^2,\quad (\epsilon=+1,~ S^6) \label{6D-positive}\\
ds\Z{6}^2 &=& dz^2+ \sinh^2{z}\, d\Omega_{5}^2,\quad (\epsilon=-1,
~H^6). \label{6D-negative}\nonumber
\end{eqnarray}
where $d\Omega_{5}^2$ is the metric of a five-sphere. This yields
${R}_{mn}=\epsilon (m-1) {g}_{mn}$. In this case, the internal
curvature has no free parameter that can be tuned or fixed
according to the choice of $\alpha$ in the warp factor. Second,
only some specific values of $\alpha$ were considered.

Although a choice as $\alpha=2$~\cite{GKP,KKLT} has often been
motivated from earlier works by Klebanov and his
collaborators~\cite{KT,KS,Zayas:2000}, in the context of
string/gauge theory dualities in a supersymmetric background, we
find no particular reason to constraint the warp factor (and the
6D curvature), especially, in a general cosmological background.

Indeed, it is not difficult to see, even with a canonical choice
of the metric as equation~(\ref{6D-positive}), that certain values
of $\alpha$ (in the warp factor) allow a de Sitter solution to
occur. In the above case, one has
\begin{equation} {}^{(10)} R_{\mu\nu}(x,z) =
{}^{(4)}\hat{R}_{\mu\nu}(x)- {\hat{g}_{\mu\nu}}\,
e^{(2+\alpha)A(z)} \left(2(2-\alpha) {A^\prime}^2 + \nabla_z^2
A\right).
\end{equation}
Thus, for some background solution with ${}^{(10)}R=0$, one gets
\begin{equation}
\nabla_z^2 e^{(2+\alpha)A}=
\frac{(2+\alpha)}{4}\,{}^{(4)}R(\hat{g})-\frac{(2-3\alpha)}{2+\alpha}\,e^{-(2+\alpha)A}
\left(\partial_z e^{(2+\alpha)A}\right)^2.
\end{equation}
While the condition like $\int \nabla_z^2 e^{(2+\alpha)A}=0$ can
be satisfied in some specific examples in AdS compactification,
with suitable boundary conditions, it is quite restrictive.
Localized sources like branes and orientifold planes will violate
such a condition. We also refer to a recent
paper~\cite{Douglas:09d} for some related discussions on warp
factor dynamics .

Let us also consider the warped volume constraint
$$ \frac{1}{G_N^{\rm eff}} \sim V_6^{\text w}
\sim \int d^6{z}\,\sqrt{g\Z{6}}\,e^{(2-3\alpha)A(z)}.
$$
We may use this relation to fix an overall normalization of warp
factor such that $G_N^{\rm eff}$ yields the 4D Newton's constant
at a boundary or 4D hypersurface where $e^{A}\sim 1$.
Particularly, with a constant warp factor $e^{A(z)}\equiv
e^{A\Z{0}}$, the warped volume $V_6^{\text w} \simeq
e^{(2-3\alpha) A\Z{0}} \times {\rm Vol}(Y_6)$ becomes finite once
${\rm Vol}(Y\Z{6})$ is fixed. But in a generic situation with a
nonconstant warp factor, the warped volume can be arbitrary large.
We will return to this discussion below.

As is evident, some of the explicit solutions found in this paper
might not respect at least one of the above-mentioned assumptions
of the no-go theorem, especially, the boundary condition, $\int
\nabla_z^2 e^{(2+\alpha)A}=0$, which is typically an additional
constraint on the warp factor. For illustration, consider the
solution (\ref{10D-nonsin}), for which
\begin{eqnarray}
{}^{(10)} R_{\mu\nu}(x,z) &=& {}^{(4)}\hat{R}_{\mu\nu}(x)-
\frac{\hat{g}_{\mu\nu}}{\lambda^2\,\sinh^2(z+z_0)}\,
e^{(2+\alpha)A(z)}
\left(2(2-\alpha) {A^\prime}^2 + \nabla_z^2 A\right) \nonumber \\
&=& {}^{(4)}\hat{R}_{\mu\nu}(x)- \frac{\hat{g}_{\mu\nu}
e^{(2+\alpha)A(z)}}{\lambda^2}\nonumber \\
&{}&\times \left[ \frac{2(2-\alpha) {A^\prime}^2+
A^{\prime\prime}}{\sinh^2(z+z_0)}
+\left(\frac{10}{\sinh2(z+z_0)}-\frac{\cosh
(z+z_0)}{\sinh^3(z+z_0)}\right)\sgn(z)\,A^\prime\right]\nonumber \\
 &=& {}^{(4)}\hat{R}_{\mu\nu}(x)- \frac{\hat{g}_{\mu\nu}}{4 \lambda^2}
\left(\frac{12}{L^2} + \frac{3(2+\alpha)}{2 L^2}\,
\coth(z+z_0)\,\delta(z)\right).\label{munu-com-10D}
\end{eqnarray}
Tracing this we can see that a de Sitter solution with
$\hat{R}^{(4)}> 0$ is possible in our model.

\subsection{How to get a finite Newton's constant?}

Although stabilization of the extra dimensional volume has been
one of the key issues in the study of inflationary solutions both
from classical supergravities and string theory, in the literature
there are no simple constructions that give rise to a finite 4D
Planck scale as well as a stabilized volume modulus. In our view,
this is not a significant drawback in our model though one could
always hope to find a robust example with all moduli fixed.

In~\cite{GKP}, Giddings {\it et al} found interest in string
compactifications with a finite 4D Planck scale. However, their
construction does not seem to reasonably produce a finite 4D
Planck mass. To understand this, it is important to note that even
when if one starts with a six-dimensional compact manifold such as
$S^6$ or compact Calabi Yau spaces, the 4D Planck mass is not
finite on its own, especially, when the 10D metric background is
warped or non-factorizable as in our examples. With a metric
ansatz of the form~(\ref{gen-coni-4}), one has
\begin{equation}
\frac{1}{G_N^{\rm eff}} \propto V_6^{\text w} \sim \int d^6 {y}
\sqrt{g\Z{6}}\,e^{(2-3\alpha)A(y)}.
\end{equation}
For the choice made in~\cite{GKP}, i.e. $\alpha=2$, one has
\begin{equation}
V_6^{\text w} \sim \int d^6 {y} \sqrt{g\Z{6}}\,e^{-4 A(y)}.
\end{equation}
As long as the warp factor is non-constant, which is obviously the
case both in the models of string flux compactifications studied
in~\cite{GKP,KKLT} and in the above examples, the 6D warped volume
would have a nontrivial dependency on some function of $y$.

For example, consider the explicit solution (\ref{most-gen-sol}),
for which
\begin{equation}
V_6^{\text w} \sim V_5 \int dy \,(y+\epsilon)^5 \,
e^{(2-3\alpha)A(y)} \propto V_5 \int
dy\,(y+\epsilon)^{(14-\alpha)/(2+\alpha)},
\end{equation}
where $V_5$ is the physical volume of the base space $X_5$ (which
is independent of $y$) and $\epsilon$ is the resolution parameter.
In the particular case that $\alpha=2$, we get
\begin{equation}
V_6^{\text w}\sim V_5 \int_{y\Z{1}}^{y\Z{2}} (y+\epsilon)^3 dy.
\end{equation}
Apparently, if $y$ is allowed to range from $0$ to $\infty$, then
the six-dimensional warped volume is not finite. This is indeed a
familiar feature of almost every inflationary solutions arising
from string theory or warped supergravity models~\cite{KS}, for
which the radial modulus is left unfixed. This particular
behaviour of the solution is not changed (except some minor
modifications) by introducing external form fields or supergravity
fluxes~\cite{Becker:2007,GKP,KKLT}. The KKLT proposal~\cite{KKLT}
is not an exception. On a general basis, for any solution to be
viable, the six dimensional volume should be small enough
(preferably exponentially suppressed) compared to the
four-dimensional volume, at least, at late times ($t\to \infty$).

An apparent divergence of the 6D volume in the above examples may
not be something that is phenomenologically ruled out or
unpleasant. To be precise, let us first consider the solution
(\ref{gen-coni-4})-(\ref{most-gen-sol}), which can be written as
\begin{equation}\label{sol-in-y}
ds_{10}^2= e^{2A(y)} \left(-dt^2 + e^{2t/{\lambda L}}\,d{\bf
x}_3^2 + \frac{32\lambda^2
L^2}{3(2+\alpha)^2}\,\frac{dy^2}{(y+\epsilon)^2} +
\frac{4\lambda^2 L^2}{3}\, ds\Z{X_5}^2\right).
\end{equation}
All of the dimensions can be small and symmetric at $t=0$. With
the expansion of the usual $3+1$ spacetime, which is exponential
in our example, the extra dimensional volume gets suppressed at
late times, $t\to \infty$. Particularly, to a 4D observer, who
uses the coordinates $x^\mu$ to measure rods and clocks, the 6D
volume is given by
\begin{eqnarray}
V_6 &=&  \frac{64\sqrt{2}}{27}\frac{\lambda^6 L^6}{(2+\alpha)}
\,V_5 \int
\frac{dy}{(y+\epsilon)}\nonumber \\
&=& \frac{2048\sqrt{2}}{729}\,\frac{\pi^3 \lambda^6
L^6}{|(2+\alpha)|}\,\ln (y+\epsilon).
\end{eqnarray}
In the above we have used the result $V_5\equiv (1/108) \int
d(\cos\theta_1) d(\cos\theta_2) d\phi_1 d\phi_2 d\psi =
16\pi^3/27$. An obvious drawback of the solution~(\ref{sol-in-y})
is that the metric is singular at $y=-\epsilon$. This is just a
coordinate artifact, indicating that $y$ is not a globally defined
coordinate.

To get physical results, including a finite 4D Planck mass, one
may introduce some elements of the Randall-Sundrum type braneworld
models~\cite{RS}. In this approach one first writes the metric
solution in terms of $z$ coordinate which may be related to the
usual conifold coordinate $y$ using the relations such $y \propto
e^{- \beta z}$ (in the $\alpha=-2$ case) or $y\propto \cosh {z}$
(in the $\alpha\ne -2$ case), or directly consider line elements
of the form (\ref{asymptote1}) or (\ref{10D-nonsin}). Then by
imposing a $Z_2$ symmetry around the brane's position at $|z|=0$,
one can show that the extra dimensional volume is effectively
finite.

As a physically more appealing example, we consider the
non-singular solution (\ref{nonsin-sols}), along with
(\ref{10D-nonsin}) and (\ref{6D-nonsin}). The 10D metric takes the
form
\begin{eqnarray}
ds\Z{10}^2 &=& e^{2A(z)} \left(-dt^2 + e^{2t/{\lambda L}}\, d{\bf
x}_3^2+
e^{-(2+\alpha)A}\lambda^2 \,ds\Z{6}^2\right)\nonumber \\
&=& e^{-\chi}\, \left(\frac{3(2+\alpha)^2 \cosh^2{(z+z_0)}}{32
L^2}\right)^{2/(2+\alpha)} \nonumber \\
&{}& \quad \times \left(-dt^2 +e^{2 t/{\lambda L}}\, d{\bf x}_3^2
+ \frac{32L^2\lambda^2}{3(2+\alpha)^2} \tanh^2{(z+z_0)} \,dz^2+
\frac{4 L^2\lambda^2}{3} \,ds\Z{X_5}^2\right),\label{non-singu5}
\end{eqnarray}
where $e^\chi$ is arbitrary. Here, without loss of generality, we
may set $z_0=0$, which corresponds to a rescaling of the radial
coordinate $z$. We thus obtain
\begin{eqnarray}\label{finite-6D}
ds_{10}^2 &=& e^{-\chi}\, \left(\frac{3(2+\alpha)^2 \cosh^2{z}}{32
L^2}\right)^{2/(2+\alpha)} \nonumber \\
&{}& \quad \times \left[ -dt^2 +e^{2 t/{\lambda L}}\, d{\bf x}_3^2
+ \frac{32 L^2 \lambda^2}{3(2+\alpha)^2} \tanh^2{z} \left(dz^2+
\frac{(2+\alpha)^2}{8} \frac{\cosh^2
{z}}{\sinh^2{z}}\,ds\Z{X_5}^2\right)\right].\nonumber\\
\end{eqnarray}
For this solution, as the result~(\ref{Kretschmann}) also implied,
both the 6D and the 10D curvature tensors are regular everywhere.
If required, by choosing $\alpha$ appropriately, such as, $\alpha<
2\sqrt{2}-2$, we can maintain a positively curved 6D manifold or a
squashed six-sphere which is topologically compact. The radial
modulus, which scales as $|\tanh z|$, is finite in the limit
$|z|\to \infty$. To a 4D observer, who uses the coordinates
$x^\mu$ to measure rods and clocks, the 6D volume is
\begin{equation}
V_6 =  \frac{2048\sqrt{2}}{729}\,\frac{\pi^3 \lambda^6
L^6}{|(2+\alpha)|}\,\ln \cosh{z}.
\end{equation}
Presumably, there is no need to make any unnatural cutoff in the
$z$ coordinate. That is, as with the usual three non-compact
directions ($x_1, x_2, x_3$), the range of the $z$ coordinate may
be chosen to be $0 \le z \le \infty$. It is interesting to note
that, with an exponential expansion of the $3+1$ spacetime, the 4D
volume becomes infinitely large at late times ($t\to \infty$),
while the extra dimensional volume grows only linearly with $z$.

\section{Effect of supergravity fluxes}

Inflationary de Sitter solutions of the above type can be obtained
in a more general class of warped compactifications or 10D
supergravities by introducing a five-form flux of the form
\begin{equation}
\tilde{F}_{(5)}=(1+*) dF(y) \wedge \Omega_4,
\end{equation}
where $F(y)$ is some function on $Y_6$ and $\Omega_4$ denotes a
volume form for $R^{3,1}$. In the case $X_5\equiv T^{1,1}$, or
$X_5=S^2\times S^3$, one can also introduce a combined three-form
flux~\cite{GKP}
\begin{equation}
G_{(3)}= F_{(3)}-\tau H_{(3)}.
\end{equation}
The 10D axion $\tau$ is generally allowed to vary over the compact
manifold, $\tau\equiv \tau (y)$.

Let us assume that the 10D metric spacetime takes the following
convenient form
\begin{equation}\label{10D-metric4}
ds\Z{10}^2 = e^{2A(y)} \hat{g}_{\mu\nu} dx^\mu dx^\nu + e^{-\alpha
A(y)}\,g_{mn}(y) dy^m dy^n.
\end{equation}
The metric tensor $g_{mn}(y)$ is arbitrary in the sense that the
internal compact space $Y_6$ has positive, zero, or negative Ricci
curvature scalar. The noncompact components of 10D Einstein
equations then take the form~\cite{GKP}
\begin{equation}\label{10D-Ricci-2}
{}^{(10)} R_{\mu\nu}= - {\hat g}_{\mu\nu} \left(
\frac{G_{(3)}^2}{48 {\text Im}\tau} + \frac{e^{-2(2+\alpha)A}}{4}
\left(\partial F\right)^2 \right) + \frac{1}{M_{10}^2}
\left(T_{\mu\nu}^{\text loc}-\frac{1}{8} {g}_{\mu\nu} T^{\text
loc}\right).
\end{equation}
With the metric ansatz (\ref{10D-metric4}), the $\mu\nu$
components of the 10D Ricci tensor read
\begin{eqnarray}
{}^{(10)}R_{\mu\nu} &=& {}^{(4)} R_{\mu\nu}(\hat{g})
-\hat{g}_{\mu\nu} e^{(2+\alpha)A}\left(\nabla_y^2 A + 2(2-\alpha)
\left(\partial_y A\right)^2 \right)\nonumber \\
&=& {}^{(4)} R_{\mu\nu}(\hat{g})-\frac{\hat{g}_{\mu\nu}}{2+\alpha}
\left(\nabla_y^2 e^{(2+\alpha)A}+ \frac{(2-3\alpha)}{(2+\alpha)}\,
\e^{-(2+\alpha)A} (\partial_y e^{(2+\alpha)A})^2 \right).
\end{eqnarray}
Using this and tracing (\ref{10D-Ricci-2}) we find
\begin{eqnarray}\label{flux-constr1}
\nabla_y^2 A &=& \frac{1}{4} \,{}^{(4)} R({\hat g})
e^{-(2+\alpha)A}- 2(2-\alpha) ({\partial_y A})^2 \nonumber\\
&{}& \quad +\, e^{-\alpha A} \frac{G_{(3)}^2}{48 {\text Im}\tau} +
\frac{e^{-(4+3\alpha) A}}{4} \left(\partial_y F\right)^2
+\frac{e^{-\alpha A}}{8
M_{10}^2}\left(T_m^m-T_\mu^\mu\right)_{\text loc}.
\end{eqnarray}
or
\begin{eqnarray}
\nabla_y^2 e^{(2+\alpha)A} &=& \frac{(2+\alpha)}{4} \,{}^{(4)}
R({\hat g})- \frac{(2-3\alpha)}{(2+\alpha)} e^{-(2+\alpha)A}
({\partial_y} e^{(2+\alpha)A})^2 \nonumber\\
&{}& + \frac{(2+\alpha)\,e^{2 A}}{4} \left(\frac{G_{(3)}^2}{48
{\text Im}\tau} + \frac{e^{-2(2+\alpha) A}}{4} \left(\partial_y
F\right)^2\right) +\frac{(2+\alpha)}{8
M_{10}^2}\left(T_m^m-T_\mu^\mu\right)_{\text loc}.\nonumber \label{flux-constr2}\\
\end{eqnarray}
The choice made by the authors of~\cite{GKP}, i.e. $\alpha=2$, is
special for which the second term on the right-hand side of
(\ref{flux-constr1}) vanishes. It is quite clear that, with a
suitable choice of $\alpha$, the 10D supergravity equations would
allow a de Sitter solution (${}^{(4)}R>0$) with nonconstant warp
factor even when the contribution from the second line in equation
(\ref{flux-constr1}) or (\ref{flux-constr2}) vanishes. The
argument in~\cite{GKP} that in the absence of localized brane
sources the fluxes must be constant and the warp factor must also
be constant does not hold, for instance, when $\alpha<2/3$ even if
$\int \nabla_y^2 e^{(2+\alpha)A}=0$. A detailed discussion about
the effects of fluxes, which is beyond the scope of this paper,
will appear elsewhere.

\section{Conclusion}

We conclude the paper with a short summary of the results. We have
presented new cosmological solutions which use generalized 6D
Einstein spaces having positive, negative or zero scalar
curvature. Our novel observation is that inflationary de Sitter
solutions can arise with all three possibilities for the internal
space curvature (i.e. $\tilde{R}_{(6)}=0$, $>0$ or $<0$), provided
one would consider the 10D metric in a sufficiently general form.

A pertinent question to ask is: Can the 4D effective cosmological
constant $\Lambda\Z{4}$ found in the above examples be tuned to be
the present value of dark energy $\Lambda\Z{0} \sim 10^{-120}
\,M\Z{\rm Pl}^2$? The answer to this question seems affirmative.
In fact, all our solutions discussed in this paper are invariant
under a constant rescaling $g_{mn}\to e^{-2\Omega}\,g_{mn}$. As a
result, there can appear an arbitrary coefficient multiplying the
10D metric spacetime. For example, consider the following 10D
metric solution
\begin{equation}\label{10D-soln3}
ds\Z{10}^2 = e^{-2{\beta}|z| -\chi} \left( -dt^2+
\exp\left(2\sqrt{\frac{4}{3}}\,e^{\Omega}\,t\right) \,d{\bf
x}\Z{3}^2 + e^{-2\Omega} \left( 2 {\beta}^2 {dz}^2 + ds\Z{5}^2
\right)\right).
\end{equation}
A constant warp factor like $e^{-\chi}$ in the above example does
not change the cosmological behaviour of our solutions, but it may
affect the value of the effective 4D Planck mass. To be more
specific, we may consider the following dimensionally reduced
action
\begin{eqnarray}
ds_{10}^2 &=& \frac{M_{10}^8}{2(2\pi)^6} \int d^{10} x
\sqrt{-g_{10}}
\left(R_{(10)}+ \cdots \right)\nonumber \\
&=& \frac{M\Z{10}^8}{\pi^3}\times \frac{\sqrt{2}}{432}\,
e^{-4\chi-6\Omega} \int d^4 x \sqrt{-g_4} \left(\hat{R}_{(4)}
-2\Lambda_4 +\cdots \right),
\end{eqnarray}
where $\Lambda_4\equiv 8 e^{2\Omega}$. We have assumed the
existence of some sort of ``brane" at the $z=0$ boundary of
spacetime, so $\int dz\,e^{-8\beta |z|}\sim
\frac{1}{4\beta}\,{\cal O}(1)$. The parameters like $e^{-\chi}$
and $e^{\Omega}$ may be fixed using some phenomenological
constraints. As a simple possibility, we may demand that
$\Lambda_4 = 8 e^{2\Omega} \sim 10^{-120}\,M_{\text Pl}^2$ and
$M_{10} > {\text TeV} \sim 10^{-15}\,M_{\text Pl}$. Then we would
have to take $\chi\gtrsim 136$. If we do not require $\Lambda_4$
to be close to present value of dark energy density, then the
parameters like $e^{\chi}$ and $e^{\Omega}$ do not have to be
fine-tuned precisely.

In our construction, no energy conditions are violated in the 4D
effective theory except the strong energy condition (SEC)
$\hat{R}^{(4)}_{tt}
>0$, or equivalently $\rho +3 p>0$, which applies only to ordinary
matter. The SEC is violated anyway by a cosmological expansion
satisfying $\hat{R}^{(4)}_{tt}\equiv -\frac{3\ddot{a}}{a}<0$ and
it may also be violated in spacetime dimensions $D <10$. In the
full 10D theory none of the energy conditions are violated (see
also~\cite{Ish:09c}).

Finally, we make a couple of remarks. In our view, for any
cosmological model to be viable in large volume limits of some
consistent string theory or supergravity compactifications, the
solutions to 10D Einstein equations are required to be nontrivial
(or cosmologically relevant) even when the external fluxes and or
non-perturbative effects are absent. The present paper meets this
requirement. Although in some specific models, for example, for
the S-brane solutions derived in~\cite{Chen:02yq}, the vacuum
solution may not be obtained just by taking a zero-flux limit of
some generic solutions, by shifting certain parameters in the
solutions~\cite{Ohta:2003pu} or by writing a general ansatz for
p-form gauge field strengths, consistent with the symmetries of
the internal manifold, the zero flux case can be treated
uniformly.

In string theory, certain background fluxes are known to be
required to satisfy certain quantization conditions~\cite{GKP},
especially, in a supersymmetric background, and also for some
other good physical and mathematical reasons, such as, for fixing
complex structure moduli associated with non-compact Calabi-Yau
spaces~\cite{KKLT}. These often place strong constraints on the
local geometry of $Y\Z{6}$ and also on the warp factor. After all,
supersymmetry is broken in our universe and in the large volume
limit, when the back reaction of the fluxes on Einstein's
equations can be ignored (since their contribution to the stress
tensor or the effective potential is volume suppressed), the warp
factor may be determined purely by 10D Einstein equations alone,
i.e. supergravity equations with zero flux. That is to say, any
consistent solution of 10D supergravity or string theory might
survive in the limit the external fluxes are turned off, giving
rise to a smooth Einstein limit. One reason behind this
expectation is that the effect of internal space curvature scales
as $e^{-(2+\alpha)A}$ while that the flux potential scales as
$e^{-2(2+\alpha)A} \int\Z{X} F^2$ ($X$ is the internal manifold
and $F$ is the field strength). Furthermore, in the large volume
limit, all possible non-perturbative effects, worldsheet $\alpha'$
corrections and string loop corrections can be negligibly small.

In some ways our models look similar to Randall-Sundrum-type
braneworld constructions with the important difference that the
effect of a 5D compact space or base manifold $X_5$ have already
been incorporated in the simple exact classical gravity solutions.
It might be possible to design string theory examples of
inflationary cosmologies by generalizing the construction in our
paper. It is also quite plausible that Ricci non-flat warped
spaces add richness to string cosmology and may potentially lead
to the realization of new cosmological scenarios.

\medskip

\noindent {\bf Acknowledgements}

\noindent I would like to thank Robert Brandenberger, Kei-ichi
Maeda, Juan Maldacena, Shinji Mukohyama, Nobu Ohta, M. Sami and
Paul Steinhardt for helpful discussions and comments. This work
was supported by the New Zealand Foundation for Research, Science
and Technology Grant no E5229, Elizabeth EE Dalton Grant no 5393
and also by the Marsden fund of the Royal Society of New Zealand.



\begin{thebibliography}{99}
\itemsep 0pt

\bibitem{supernovae}
A. G. Riess {\it et al} (Supernova Search Team Collaboration),
  Astron.\ J.\  {\bf 116}, 1009 (1998);
  S. Perlmutter {\it et al}
(Supernova Cosmology Project Collaboration), Astrophys. J. {\bf
517}, 565 (1999).

\bibitem{WMAP}
D. N. Spergel {\it et al} (WMAP Collaboration),
Astrophysical  Journal Suppl. 148 (2003) 175;
{\it ibid}, {\bf 170}, 377 (2007).

\bibitem{KKLT}
S.~Kachru, R.~Kallosh, A.~Linde and S.~P.~Trivedi,
  {\it De Sitter vacua in string theory},
  Phys.\ Rev.\  D {\bf 68}, 046005 (2003);
%
C.~P.~Burgess, J.~M.~Cline, H.~Stoica and F.~Quevedo,
  {\it Inflation in realistic D-brane models},
  JHEP {\bf 0409} (2004) 033.

\bibitem{Ish03c}
I.~P.~Neupane,
  {\it Accelerating cosmologies from exponential potentials},
  Class.\ Quant.\ Grav.\  {\bf 21}, 4383 (2004);
N.~Ohta,
  {\it Accelerating cosmologies and inflation from M / superstring
  theories}m
  Int.\ J.\ Mod.\ Phys.\  A {\bf 20} (2005) 1
  [arXiv:hep-th/0411230].



\bibitem{Becker:2007}
  M.~Becker, L.~Leblond and S.~E.~Shandera,
  {\it Inflation from Wrapped Branes},
  Phys.\ Rev.\  D {\bf 76} (2007) 123516.


\bibitem{Chen:02yq}
  C.~M.~Chen, D.~V.~Gal'tsov and M.~Gutperle,
  Phys.\ Rev.\  D {\bf 66}, 024043 (2002)
  [arXiv:hep-th/0204071].

\bibitem{Ohta:2003pu}
  P.~K.~Townsend and M.~N.~R.~Wohlfarth,
  {\it Accelerating cosmologies from compactification},
  Phys.\ Rev.\ Lett.\  {\bf 91} (2003) 061302;
N.~Ohta,
  {\it Accelerating cosmologies from S-branes},
  Phys.\ Rev.\ Lett.\  {\bf 91} (2000) 061303;
{\it A study of accelerating cosmologies from superstring / M
theories},
  Prog.\ Theor.\ Phys.\  {\bf 110} (2003) 269.

\bibitem{Ish03a}
  C.~M.~Chen, P.~M.~Ho, I.~P.~Neupane, N.~Ohta and J.~E.~Wang,
  {\it Hyperbolic space cosmologies},
  JHEP {\bf 0310} (2003) 058 [arXiv:hep-th/0306291].
%
\bibitem{Ish05a}
I.~P.~Neupane and D.~L.~Wiltshire,
  {\it Accelerating cosmologies from compactification with a
  twist},
  Phys.\ Lett.\  B {\bf 619} (2005) 201;
  {\it Cosmic acceleration from M theory on twisted spaces},
  Phys.\ Rev.\  D {\bf 72} (2005) 083509.

\bibitem{Ish07a}
I.~P.~Neupane,
  {\it Accelerating universes from compactification on a warped
  conifold},
  Phys.\ Rev.\ Lett.\  {\bf 98} (2007) 061301.

\bibitem{GKP}
  S.~B.~Giddings, S.~Kachru and J.~Polchinski,
  {\it Hierarchies from fluxes in string compactifications},
  Phys.\ Rev.\  D {\bf 66} (2002) 106006.

\bibitem{RS}
L.~Randall and R.~Sundrum, {\it A large mass hierarchy from a
small extra dimension}, Phys.\ Rev.\ Lett.\ {\bf 83} (1999) 3370;
%
  {\it An alternative to compactification},
{\it ibid}, Phys.\ Rev.\ Lett.\  {\bf 83} (1999) 4690.

\bibitem{KT}
I.~R.~Klebanov and A.~A.~Tseytlin,
  {\it Gravity Duals of Supersymmetric SU(N) x SU(N+M) Gauge
  Theories},
  Nucl.\ Phys.\  B {\bf 578} (2000) 123.

\bibitem{Gibbons-84}
G.~W.~Gibbons, in {\it Supersymmetry, Supergravity and Related
Tolics}, edited by F. del Aguila, J. A. de Azcarraga, and L. E.
Ibanz (World Scientific, Singapore, 1985), pp. 123-146.


\bibitem{Malda:2000}
  J.~M.~Maldacena and C.~Nu\~nez,
  {\it Supergravity description of field theories on curved manifolds and a no  go
  theorem},
  Int.\ J.\ Mod.\ Phys.\  A {\bf 16} (2001) 822.


\bibitem{Zayas:2000}
  L.~A.~Pando Zayas and A.~A.~Tseytlin,
  {\it 3-branes on resolved conifold},
  JHEP {\bf 0011} (2000) 028 [arXiv:hep-th/0010088].


\bibitem{KS}
I.~R.~Klebanov and M.~J.~Strassler,
  {\it Supergravity and a confining gauge theory: Duality cascades and
  $\chi$-SB-resolution of naked singularities},
  JHEP {\bf 0008} (2000) 052.


\bibitem{Wesley}
D.~H.~Wesley,
  {\it Oxidised cosmic acceleration},
  JCAP {\bf 0901} (2009) 041.

\bibitem{Douglas:09d}
  M.~R.~Douglas,
  arXiv:0911.3378 [hep-th].

\bibitem{Ish:09c}
  I.~P.~Neupane,
  {\it Extra dimensions, warped compactifications and cosmic
  acceleration}, Phys. Lett. B 683 (2010) 88
  [arXiv:0903.4190];
{\it Accelerating universe from warped extra dimensions},
  Class.\ Quant.\ Grav.\  {\bf 26} (2009) 195008
  [arXiv:0905.2774].








\end{thebibliography}
\end{document}